\documentclass[preprint,aps,showpacs,nofootinbib]{revtex4}
\usepackage{bm,amssymb,amsmath,cancel,footnote,subfigure}
 \usepackage{feynmp}
 \usepackage[pdftex]{graphicx}
\begin{document}

\count255=\time\divide\count255 by 60 \xdef\hourmin{\number\count255}
  \multiply\count255 by-60\advance\count255 by\time
 \xdef\hourmin{\hourmin:\ifnum\count255<10 0\fi\the\count255}

\newcommand{\xbf}[1]{\mbox{\boldmath $ #1 $}}
\newcommand{\Dslash}{D\hspace{-0.7em}{ }\slash\hspace{0.2em}}

\title{Large-$N$ Structure of Tetraquark Mesons}

%

\author{Richard F. Lebed}
\email{Richard.Lebed@asu.edu}

\affiliation{Department of Physics, Arizona State University, 
Tempe, AZ 85287-1504}


\date{August 2013}

\begin{abstract}
Following Weinberg's argument that narrow tetraquark mesons are not
precluded in large $N_c$ QCD, we explore the flow of $N_c$ factors
needed for the consistency of this picture, and show that they must
arise in a novel way, not simply through the usual combinatoric
counting.
\end{abstract}


\maketitle
\thispagestyle{empty}

\newpage
\setcounter{page}{1}


Recently, Weinberg~\cite{Weinberg:2013cfa} re-evaluated the well-known
argument of Coleman~\cite{Coleman}, that color-singlet operators of
the form $q\bar q q\bar q$ create nothing but meson pairs, to propose
that such operators can consistently create narrow tetraquarks [widths
of $O(1/N_c)$, the same as ordinary mesons] in the limit of a large
number $N_c$ of QCD colors.  Subsequently, Knecht and
Peris~\cite{Knecht:2013yqa} showed that some tetraquarks of this type
would mix with mesons at $O(N_c^0)$ and hence be very difficult to
disentangle in experimental data, but they also showed that
tetraquarks of certain flavor structures do not suffer this fate and
may be discernible in data.  Our purpose in this short note is to show
that narrow tetraquark states, if they indeed exist, exhibit an $N_c$
behavior at the quark level that is novel compared to that of
conventional meson or baryon states.

In either approach, one begins with the color-singlet quartic quark
operator
\begin{equation}
{\cal Q}(x) = \sum_{ij} C_{ij} {\cal B}_i (x) {\cal B}_j (x) \, ,
\label{eq:quartic}
\end{equation}
where ${\cal B}_i(x)$ are color-singlet quark bilinears,
\begin{equation}
{\cal B}_i (x) = \sum_{a = 1}^{N_c} \overline{q^a} (x) \Gamma_i \,
q^a(x) \, ,
\label{eq:bilinear}
\end{equation}
which (if needed) may be redefined to have zero vacuum expectation
values, the index $a$ sums over all $N_c$ colors, $\Gamma^i$
represents the $N_c$-independent spin-flavor connections defining
${\cal B}_i$, and $C_{ij} = C_{ji}$ are a collection of numerical
coefficients (taken $N_c$-independent in Ref.~\cite{Weinberg:2013cfa})
that describe the correlation of the bilinears ${\cal B}_i$, ${\cal
  B}_j$ at the spacetime point $x$.  It is a feature of SU($N_c$)
group theory (operationally manifested by Fierz reordering) that any
color-singlet quartic ${\cal Q}$ can be decomposed into such a
product.  The interesting quantity to study is the ${\cal Q}$
two-point function, which has disconnected and connected pieces:
\begin{eqnarray}
\langle {\cal Q} (x) {\cal Q}^\dagger (y) \rangle & = &
\sum_{ijkl} C_{ij} C_{kl} \left[ \langle {\cal B}^{\vphantom\dagger}_i
(x) {\cal B}^\dagger_k (y) \rangle \langle
{\cal B}^{\vphantom\dagger}_j (x) {\cal B}^\dagger_l (y) \rangle
+
\langle {\cal B}^{\vphantom\dagger}_i(x) {\cal B}^{\vphantom\dagger}_j
(x) {\cal B}^\dagger_k (y) {\cal B}^\dagger_l (y) \rangle_{{\rm conn}}
\right] .
\label{eq:twopoint}
\end{eqnarray}
Suppose that Eq.~(\ref{eq:quartic}) is generalized to a nonlocal form
by allowing the spacetime point $x$ appearing in the two bilinears to
separate into distinct values $x_1 \neq x_2$, and that such bilinears
are used in the definition of Eq.~(\ref{eq:twopoint}) with $y \to y_1
\neq y_2$ as well.  The coefficients in this case generalize to
$C_{ij} (x_1 - x_2) C_{kl} (y_1 - y_2)$.  Then, by conventional large
$N_c$ counting, the first term is $O(N_c^2)$ and is saturated by sums
over the propagators of two noninteracting mesons, each of which (by
LSZ reduction) has a decay constant $f_{\cal B} = O(N_c^{1/2})$
appearing for each asymptotic state, thus neatly accounting for the
leading-order $N_c$ behavior.  The second term is $O(N_c^1)$ and
represents, at least in part, two-meson scattering diagrams; once the
four $f_{\cal B}$ factors are amputated, one finds the well-known
result that meson-meson scattering amplitudes are only $O(1/N_c)$.

Now allow the points $x_{1,2}$ and $y_{1,2}$ to again coalesce to $x$
and $y$, respectively, as in Eq.~(\ref{eq:quartic}).  Then the $N_c$
counting remains the same as before, but poles due to localized
one-tetraquark states may now be considered.  However, true
propagating tetraquark states can only appear in the connected term,
which is subleading in $1/N_c$.  From this fact, Ref.~\cite{Coleman}
concludes that the two-point function of Eq.~(\ref{eq:twopoint})
produces only two-meson states, but Ref.~\cite{Weinberg:2013cfa}
concludes that nothing precludes the production of tetraquark states
in its subleading, connected piece.  Moreover, since this $O(N_c^1)$
term contains the full amplitude for the free $O(N_c^0)$ propagation
of a tetraquark, LSZ reduction shows that the tetraquark decay
constant $f_{\cal Q} = O(N_c^{1/2})$, and by using this result on the
$O(N_c^1)$ connected three-point function $\langle {\cal Q} (x) {\cal
  B}_i (y) {\cal B}_j (z) \rangle_{\rm conn}$, one finds the trilinear
coupling to be $O(N_c^{-1/2})$, and hence the tetraquark decay width
to be $O(1/N_c)$.  These are the central results of
Ref.~\cite{Weinberg:2013cfa}.

The point of interest in this note is the interpretation of the
tetraquark decay constant $f_{\cal Q}$ in terms of the underlying
quark degrees of freedom.  In the case of mesons, the fact that
$f_{\cal B} = O(N_c^{1/2})$ is equivalent to the fact that the LSZ
reduction identifies the operator ${\cal B}/\sqrt{N_c}$ as the one
that creates or destroys a properly normalized meson state.  Viewed in
terms of the quark color degrees of freedom in
Eq.~(\ref{eq:bilinear}), this prefactor is simply the one that
produces a normalized wave function over the color degrees of freedom
$c_a$, $\sum_{a=1}^{N_c} \bar c_a c_a / \sqrt{N_c}$.  In the case of
the tetraquark, however, Eq.~(\ref{eq:twopoint}) indicates that the
color wave function of the operator ${\cal Q}$ has a norm-square of
$N_c^2$, and that of the operator ${\cal Q}/\sqrt{N_c}$, which
according to Ref.~\cite{Weinberg:2013cfa} creates or destroys
normalized tetraquark states, actually has norm-square of $N_c^1$
rather than unity.  How can these facts be made to agree?

In fact, while the operator ${\cal Q}/\sqrt{N_c}$ creates and destroys
normalized tetraquarks, it need not produce tetraquarks alone.
Indeed, our argument that the connected nonlocal four-bilinear
correlation function is dominated by two-meson scattering suggests
that a dominant component of the ${\cal Q}/\sqrt{N_c}$ two-point
function is represented by coincident (propagating from $y$ to $x$)
pairs of mesons, including meson molecules.  The authors of
Refs.~\cite{Weinberg:2013cfa,Knecht:2013yqa} assert that narrow
tetraquarks {\em can\/} exist; our purpose is to find a condition that
causes them to arise naturally.  In order to obtain the additional
$1/\sqrt{N_c}$ suppression for creating properly normalized
tetraquarks, we note that pointlike tetraquark states appear only in
the local limits $x_2 \to x_1$, $y_2 \to y_1$, or at least when $x_1 -
x_2$ and $y_1 - y_2$ are sufficiently small that the tetraquark wave
function has appreciable amplitude at all four of these points; but
then, the use of LSZ reduction (which requires well-separated sources)
to identify four distinct meson sources is no longer justified.  One
may, for example, model the required behavior by allowing the
coefficients $C_{ij}$ of Eq.~(\ref{eq:quartic}) in the nonlocal case
to have a contribution of the Gaussian form
\begin{equation}
\delta C_{ij} \sim \exp [ -N_c^m \Lambda_{\rm QCD}^2 (x_1 - x_2)^2 ]
\, ,
\label{eq:profile}
\end{equation}
where $m > 0$.  For any finite separation $(x_1 -x_2)^2 \agt
1/\Lambda_{\rm QCD}^2$, one finds $\delta C_{ij} \to 0$ as $N_c \to
\infty$, so that tetraquark poles are features of the connected
four-bilinear correlator only in the local limit and do not modify the
naive $O(N_c^0)$ counting of $C_{ij}$.  However, integrating $\delta
C_{ij}$ over an arbitrarily small range of separation between $x_1$
and $x_2$---that is to say, averaging over the $O(1/\Lambda_{\rm
  QCD})$ spatial extent of the tetraquark wave function---gives a
contribution of $O(N_c^{-m/2})$ for each integral taken over $C_{ij}$.
For the special case of 3 spatial integrals and $m = \frac 1 3$, one
obtains precisely the desired $1/\sqrt{N_c}$ behavior to obtain a
narrow tetraquark with $f_{\cal Q} = O(\sqrt{N_c})$.  Any particular
choice for carrying out this averaging is suitable, as long as it
integrates over the whole functional peak.

The proposed profile $\delta C_{ij}$ presented here is something both
new and familiar in large $N_c$ counting.  It arises through the
noncommutativity of the $(x_1 - x_2)^2 \to 0$ and $N_c \to \infty$
limits.  $\delta C_{ij}$ is not obtained from any first-principles
counting of factors of the QCD coupling constant $g_s \sim
1/\sqrt{N_c}$ and combinatoric $N_c$ color
factors~\cite{'tHooft:1973jz}, which are sufficient to explain the
basic $N_c$ counting for ordinary mesons and glueballs, hybrid
mesons~\cite{Cohen:1998jb}, baryons~\cite{Witten:1979kh}, and baryon
resonances~\cite{Cohen:2003tb}.  Indeed, $\delta C_{ij}$ is
nonperturbative in $N_c$ counting, and since $1/g_s^2 \sim N_c$, the
profile is reminiscent of an instanton structure.  Nevertheless, we do
not assert the presence of any specific instanton-like physics at work
here, but rather only the need for a contribution to $C_{ij}$ with
distinct local and nonlocal limits with respect to $N_c$; any
functional form that produces the $1/\sqrt{N_c}$ for $(x_1 - x_2)^2
\to 0$ is suitable for this purpose.  Finally, the appearance of a
$1/\sqrt{N_c}$-suppressed four-quark state arising in the limit in
which two ${\cal B}$ operators converge in the $O(N_c^1)$ three-point
function $\langle {\cal B}_i (x) {\cal B}_j (y) {\cal B}_k (z)
\rangle_{\rm conn}$ provides a mechanism through which
meson-tetraquark mixing can occur at $O(N_c^0)$~\cite{Knecht:2013yqa}.

In summary, we propose a resolution to the problem of $N_c$ counting
for the proposed narrow tetraquark states of
Ref.~\cite{Weinberg:2013cfa}, and are led to a novel application of
inducing the required $N_c$ factors via a term nonperturbative in
$N_c$.

\begin{acknowledgments}
The author is indebted to S.~Weinberg and T.~Cohen for helpful
discussions.  This work was supported by the National Science
Foundation under Grant No.\ PHY-1068286.
\end{acknowledgments}

\end{document}